\documentclass[preprint,aps,prc,showpacs,nofootinbib]{revtex4}
\usepackage{epsfig}
\usepackage{graphicx}

\newcommand\ba{\begin{eqnarray}}
\newcommand\ea{\end{eqnarray}}
\newcommand\nn{\nonumber}

\begin{document}
\title{Testing the RRPP vertex of effective Regge action}

\author{E. A. Kuraev, V. V. Bytev, S. Bakmaev}
\affiliation{\it JINR-BLTP, 141980 Dubna, Moscow region, Russian
Federation}

\author{E.N. Antonov}

\affiliation{\it Petersburg Nuclear physics Institute, RU-188350 PNPI, Gatchina, St. Petersburg, Russia}

\date{\today}

\begin{abstract}
In frames of effective Regge action the vertices describing
conversion of two reggeized gluons to one two and three ordinary
gluons was constructed. The self-consistency: Bose symmetry and
gauge invariance properties checks was shown to be fulfilled. The
simplest one with creation of a single gluon was intensively
verified in programs of experimental and theoretical treatment
since it determine the kernel of of the known BFKL equation. Here
we discuss the possibility to check the vertex with creation of
two real gluons, which can reveal itself in process of scalar
mesons production in high energy peripheral nucleons collisions.
We show that the mechanisms which include emission of two gluons
in the same effective vertex contribution dominate compared with
one with the creation of two separate gluons. Numerical
estimations of cross section of pair of charged pions production
for LHC facility  give the value or order $10 mb$. As well we
estimate the excess of production of positively charged muons (as
a decay of pions) created by cosmic ray proton collisions with the
atmosphere gas nuclei to be in a reasonable agreement with modern
data.
\end{abstract}

\maketitle
\section{Introduction}
The problem of unitarization of the BFKL Pomeron as a composite
state of two interacting reggeized gluons \cite{Kuraev} is actual
problem of QCD \cite{Lipatov}. In the circle of questions are the
construction of three Pomeron vertices, and more complicated ones
including as well the ordinary gluons can be solved using in terms
of vertices of effective Regge action formulated in form of
Feynman rules for effective Regge action build in \cite{Antonov}.
In this paper the effective vertices containing the ordinary and
the reggeized gluons were build.

A lot of attention to describe the creation in the peripheral
kinematics of the bound states of heavy and light quarks was paid
recent time \cite{Saleev}. Creation of gluon bound states
(gluonium) in the peripheral kinematics was poorly considered in
literature. This is a motivation of this paper.

To satisfy the requirement for the scattered nucleons to be colorless states (non excited barions), we must consider
the Feynman amplitudes corresponding to two reggeized gluons exchange. More over in each nucleon both reggeized gluons
must interact with the same quark.

Process of creation of two gluons with the subsequent conversion
of them to the bound state with quantum number of $\sigma$ meson
in the pionization region of kinematics of nucleons collisions:
\ba N_A(P_A)+N_B(P_B) \to
N_A(P_A')+N_B(P_B')+g_1(p_1)+g_2(p_2);\quad g_1(p_1)+g_2(p_2)\to
\sigma(p), \ea can be realized by two different mechanisms. One of
them consist in creation of two gluons each of them is created in
"collision" of two reggeized gluons RRP - type vertice. Another
contains the effective vertex with two reggeized and two ordinary
gluons RRPP - vertex. The second reggeized gluon in the scattering
channel do not create real gluons. the relevant
$RRP$ vertex have a form: \ba V_\mu(r_1(q_1,a)+r_2(q_2,b)\to
g(\mu,c))=\Gamma^\mu_{cad}(q_1,k,q_2)=g f_{cad}C_\mu(q_1,k,q_2),
\ea with the coupling constant $g$, $g^2=4\pi\alpha_s$ and
$k=q_1+q_2$, \ba
C_\mu(q_1,k,q_2)=2[(n^-)^\mu\left(q_1^++\frac{q_1^2}{q_2^-}\right)-(n^+)^\mu\left(q_2^-+\frac{q_2^2}{q_1^+}\right)+(q_2-q_1)_\mu].
\ea The 4-vector $C_\mu$ obey the gauge condition $k^\mu C_\mu=0$.

The effective $RRPP$ vertex with the conservation law
$$r_1(q_1,c)+r_2(q_2;d)\to g_1(p_1,\nu_1,a_1)+g_2(p_2, \nu_2,a_2)$$
have a form \cite{Antonov}: \ba
\frac{1}{ig^2}\Gamma^{\nu_1\nu_2}_{ca_1a_2d}(q_1,p_1,p_2;q_2)=\frac{T_1}{p_{12}^2}c^\eta(q_1,q_2)\gamma^{\nu_1\nu_2\eta}(-p_1,-p_2,k)+ \nn \\
\frac{T_3}{(p_2-q_2)^2}\Gamma^{\eta\nu_1-}(q_1,p_1-q_1)\Gamma^{\eta\nu_2+}(p_2-q_2,q_2)- \nn \\
\frac{T_2}{(p_1-q_2)^2}\Gamma^{\eta\nu_2-}(q_1,p_2-q_1)\Gamma^{\eta\nu_1+}(p_1-q_2,q_2)-T_1[(n^-)^{\nu_1}(n^+){\nu_2}- \nn \\
(n^-)^{\nu_2}(n^+){\nu_1}]-T_2[2g^{\nu_1\nu_2}-(n^-)^{\nu_1}(n^+){\nu_2}]-T_3[(n^-)^{\nu_2}(n^+){\nu_1}-2g_{\nu_1\nu_2}]- \nn \\
2q_2^2(n^+)^{\nu_1}(n_+)^{\nu_2}[\frac{T_3}{p_2^+q_1^+}-\frac{T_2}{p_1^+q_1^+}]- \nn \\
2q_1^2(n^-)^{\nu_1}(n_-)^{\nu_2}[\frac{T_3}{p_1^-q_2^-}-\frac{T_2}{p_2^-q_2^-}],
\ea
with the light-like 4-vectors
$$n^+=p_B/E,n^-=p_A/E, n^+n^-=2,(n^\pm)^2=0,$$
$\sqrt{s}=2E$ is the total center of mass energy,
 \ba
\gamma_{\mu\nu\lambda}(p_1,p_2,p_3)=(p_1-p_2)_\lambda
g_{\mu\nu}+(p_2-p_3)_\mu g_{\nu\lambda}+(p_3-p_1)_\nu
g_{\lambda\mu},p_1+p_2+p_3=0, \ea is the ordinary three gluon
Yang-Mills vertex, and the induced vertices: \ba
\Gamma^{\nu\nu'+}(q_1,q_2)=2q_1^+g^{\nu\nu'}-(n^+)^\nu(q_1-q_2)^{\nu'}-(n^+)^{\nu'}(q_1+2q_2)^\nu-
\frac{q_2^2}{q_1^+}(n^+)^\nu(n^+)^{\nu'}; \ea \ba
\Gamma^{\nu\nu'-}(q_1,q_2)=2q_2^-g^{\nu\nu'}+(n^-)^\nu(q_1-q_2)^{\nu'}+(n^-)^{\nu'}(-q_2-2q_1)^\nu-
\frac{q_1^2}{q_2^-}(n^-)^\nu(n^-)^{\nu'}. \ea We use here the
notation $k^\pm=(n^\pm)_\mu k^\mu$ and light cone decomposition
implies: \ba
q_1=q_{1\bot}+\frac{q_1^+}{2}n^-;q_2=q_{2\bot}+\frac{q_2^-}{2}n^+; q_1^-=q_2^+=0, \nn \\
p_i=\frac{p_i^+}{2}n^-+\frac{p_i^-}{2}n^++p_{i\bot}, p_\bot
n^\pm=0. \ea The color structures are \ba
T_1=f_{a_1a_2r}f_{cdr},T_2=f_{a_2cr}f_{a_1dr};
T_3=f_{ca_1r}f_{a_2dr}, \ea with $f_{abc}$ is the structure
constant of the color group; Jacoby identity provides the relation
$T_1+T_2+T_3=0$. The conditions of Bose-symmetry and gauge
invariance: \ba
\Gamma^{\nu_1\nu_2}_{ca_1a_2d}(q_1,p_1,p_2,q_2)p_{1\nu_1}=0,\Gamma^{\nu_1\nu_2}_{ca_1a_2d}(q_1,p_1,p_2,q_2)=
\Gamma^{\nu_2\nu_1}_{ca_2a_1d}(q_1,p_1,p_2,q_2), \ea are
satisfied.

\section{Pomeron mechanisms of $\sigma$ meson production}

We consider the case when the hadrons after collision remains to be colorless. For the case of nucleon collisions
it results that both exchanged reggeized gluons must interact with the same quark.
The color coefficient associated with $RRPP$ vertex results to be
\ba
Tr t^n t^c \times Tr t^n t^d\times\Gamma_{ca_1a_2d}^{\nu_1\nu_2}=\frac{1}{4}N \delta_{a_1a_2}\Pi_{02}^{\nu_1\nu_2},
\ea
$N=3$ is the rank of the color group.
For the mechanism of creation of two separate gluons we have
\ba
Tr t^n t^k \times Tr t^m t^l \times f_{nma_1}\times f_{kla_2}\times\Pi_{11}^{\nu_1\nu_2}=
\frac{1}{4}N \delta_{a_1a_2}\Pi_{11}^{\nu_1\nu_2}.
\ea
Projecting the two gluon state to the colorless and spin-less state we use the operator:
\ba
{\cal P}=\frac{\delta_{a_1a_2}}{\sqrt{N^2-1}}\frac{g^{\nu_1\nu_2}}{4}.
\ea
The resulting expressions are
$$
\frac{1}{16}N\sqrt{N^2-1}[\Pi_{02},\Pi_{11}]
$$
with
\ba
\Pi_{02}=-12-[\frac{1}{(p_2-q_2)^2}\Gamma^{\eta\nu-}(q_1,p_1-q_1)\Gamma^{\eta\nu+}(p_2-q_2,q_2)+(p_1\leftrightarrow p_2)];
\ea
$q_1=l_1,q_2=p_1+p_2-l_1$ and
\ba
\Pi_{11}=C_\mu(l-l_1,l_1-l+p_1)C_\mu(l_1,p_1-l_1),
\ea
Here $l_1$ is the 4-momentum of the gluonic loop, $l=P_A-P_{A'}$ is the transferred momentum.

In a realistic model describing interaction two reggeized gluons
with the transversal momenta $\vec{l}_1,\vec{l}_2$, which form a
Pomeron with quark \cite{Gunion} we have for the corresponding
vertex \ba
\Phi_P(\vec{l}_1,\vec{l}_2)=-\frac{12\pi^2}{N}F_P(\vec{l}_1,\vec{l}_2),
\ea with \ba F_P(\vec{l}_1,\vec{l}_2)=\frac{-3\vec{l}_1\vec{l}_2
C^2}{(C^2+(\vec{l}_1+\vec{i}_2)^2)(C^2+\vec{l}_1^2+\vec{l}_2^2-\vec{l}_1\vec{l}_2)},
\ea and $C=m_\rho/2\approx 400 MeV$. We note that this form of
Pomeron-quark coupling obey gauge condition: it turns to zero at
zero transverse momenta of gluons. The factor 3 corresponds to
three possible choice of quark into proton.

Matrix elements of peripheral processes are proportional to $s$.
To see it we can use Gribov's substitution  to gluon Green
functions nominators $g_{\mu\nu}=(2/s)P_{A\mu}P_{B\nu}$ with
Lorentz index $\mu(\nu)$ is associated with $B(A)$ parts of
Feynman amplitude. Performing the loop momenta $l_1$ integration
it is convenient to use such a form of phase volume \ba
d^4l_1=\frac{1}{2s}ds_1ds_2d^2\vec{l}_1, s_1=2P_Al_1,s_2=2P_Bl_1.
\ea Simplifying the nucleon nominators as \ba
\bar{u}(P_A')\hat{P}_B\hat{P}_A\hat{P}_B u(P_A)=s^2N_A,
N_A=\frac{1}{s}\bar{u}(P_A')\hat{P}_B u(P_A),\sum |N_A|^2=2, \ea
(with the similar expression for part $B$), we find all they be
equal. The integration on variables $s_{1,2}$ can be performed for
the sum of all four Feynman amplitudes as \ba
\int\limits_{-\infty}^\infty\frac{ds_{1,2}}{2\pi
i}[\frac{1}{s_{1,2}+a_{1,2}+i0}+\frac{1}{-s_{1,2}+b_{1,2}+i0}]=1.
\ea

Combining all the factors the matrix element corresponding to the vertex RRPP can be written as
\ba
-iM=2^53^2s(\pi\alpha_s)^3N_A N\sqrt{N^2-1}\frac{1}{N}F(\vec{\Delta})f(\vec{l},\vec{p}),
\ea
with $\vec{p}=\vec{p}_1+\vec{p}_2$ and the relative momenta of real gluons $\vec{\Delta}=(\vec{p}_2-\vec{p}_1)/2$,
\ba
f(\vec{l},\vec{p})=\int\frac{d^2l_1C^4}{2\pi\vec{l}_1^2(\vec{l}-\vec{l}_1)^2(\vec{p}-\vec{l}_1)^2}
F_P(\vec{l}_1,\vec{l}-\vec{l}_1)F_P(\vec{l}_1-\vec{l},\vec{p}-\vec{l}_1)\Pi_{02}(\vec{l}_1, \vec{p}-\vec{l}_1),
\ea
and the similar expression for another mechanism.
Here we had introduced the factor$F(\vec{\Delta})=[a^2\vec{\Delta}^2+1]^{-2}$
which describe the conversion of two gluon state to bound state of the size $a\sim 1 fm$ which is gluonium component of scalar meson.

Let perform the phase volume of the final state as
\ba
d\Gamma=\frac{d^3P_B'}{2E_{B'}}\frac{d^3P_A'}{2E_{A'}}\frac{d^3p_1}{2E_1}\frac{d^3p_2}{2E_2}(2\pi)^{-8}
\delta^4(P_A+P_B-P_{A'}-P_{B'}-p_1-p_2)\nn \\= d\Gamma_{AB} d \Gamma_{12}(2\pi)^{-8}, \nn \\
d\Gamma_{AB}=d^4ld^4P_{B'}d^4P_{A'}\delta^4(P_A-P_{A'}-l)\times \nn \\
\times\delta^4(P_B+l-P_{B'}-p)\delta(P_{B'}^2-M_B^2)\delta(P_{A'}^2-M_A^2), \nn \\
d\Gamma_{12}=\frac{d^3p}{2E_1}\frac{d^3\Delta}{2E_2}. \ea Using
the relation \ba d^4l=\frac{1}{2s}d(2lP_B)d(2lP_A)d^2\vec{l}, \ea
we perform integration on the momenta of the scattered nucleons
with the result: \ba d\Gamma_{AB}=\frac{d^2\vec{l}}{2s}. \ea
Keeping in mind the almost collinearity of 3-momenta of real gluon
we can transform the gluon part of the phase volume as: \ba
d\Gamma_{12}=d^4p\delta(p^2-M_\sigma^2)\frac{2}{M_\sigma}d^3\Delta.
\ea Integration on the $\vec{\Delta}$ can be performed in the
explicit form: \ba \int d^3\Delta
F^2(\Delta)=\frac{\pi^2}{4a^3}=\frac{2\pi^2M_p^3}{10^3a(fm)^3},
\ea where we had used the conversion constant $M_p\times fm=5$,
$M_p$ is the nucleon mass. The last part of the phase volume of
gluonic system can be arranged using light cone form of 4-momentum
$p$ (see (8)): \ba
d^4p\delta(p^2-M_\sigma^2)=\frac{dp^+}{2p^+}d^2\vec{p}=\frac{1}{2}L
d^2\vec{p}, \ea with the so called "boost logarithm"
$L\approx\ln(\frac{2E}{M}),\quad E, M - $ Energy and mass of
proton in laboratory reference frame. For LHC facility as well as
for cosmic protons (in the knee region of spectra) we use below
$L=15.$

For the contribution to the total cross section we obtain
\ba
\sigma^P_1=A\frac{\alpha_s^6L M_p^3}{M_\sigma M_\rho^4}J,A=\frac{6^4}{5^3}\frac{N^2-1}{N^2}\frac{\pi^2}{a(fm)^3},
\ea
with
\ba
J=\int\frac{d^2ld^2p}{(2\pi)^2C^4}f^2(\vec{l},\vec{p}).
\ea
Numerical integration give $J=7.4*10^3$.

Corresponding contribution to the total cross section of the
single $\sigma$ meson production is of order of $10mb$.

The contribution arising from the other mechanism of production (including the interference
of amplitudes) turns out be at least order of magnitude less. It determines the accuracy of the result obtained
on the level of $10\%$.

\section{Screening effects. Several $\sigma$- production}

Let us now generalize the result to include the screening effects as well as the possibility to
produce several $\sigma$ mesons.

At large impact parameters limit proton interact with the whole
gluon field of the nucleon (or nuclei) moving in the opposite
direction coherently. So the main contribution arises from the
many Pomeron exchanges mechanism (compare with the "chain'
mechanism essential in BFKL equation) \cite{Kuraev}. So we must
consider Pomeron s-channel iterations. Let consider three kinds of
the iteration blocks. One is the pure Pomeron exchange, the second
is Pomeron with the sigma meson emission from the central region.
The third one is the "screening block":two blocks of the second
type with the common virtual $\sigma$-meson. Contribution to the
amplitude of production $n$ $\sigma$ mesons of the blocks of the
third kind is associated with the "large logarithm" which arises
from boost freedom of these blocks in completely analogy with QED
\cite{BGKN}.

In the similar way the closed expression (omitting the terms of order $1/N^2$ compared with the ones of order of $1$) for
the summed on number of $s$-channel iteration ladders
of the first and the third type can be obtained using the relation
\ba
\int\Pi_1^n\frac{d^2k_i}{(2\pi)^2}=\int\Pi_1^{n+1}\frac{d^2k_i}{(2\pi)^2}\int d^2\rho\exp(i\vec{\rho}\sum_1^{n+1}(\vec{k}_i-\vec{q}))= \nn \\
\int d^2\rho\exp(-i\vec{q}\vec{\rho})\Pi_1^{n+1}\frac{d^2k_i}{(2\pi)^2}e^{i\vec{k}_i\vec{\rho}}.
\ea
Accepting the assumption about color-less structure of the Pomeron as a bound state of two reggeized interacting gluons
and applying the same sequence of transformations as was done in \cite{BGKN} we obtain for the cross section of $n$ $\sigma$
mesons production at the peripheral high energy protons collisions:
\ba
\sigma_n=\int\frac{d^2\rho}{(2\pi)^2}\frac{(L\sigma_0Z(\rho))^n}{n!}exp(-L\sigma_0 Z(\rho)),
\ea
with $\sigma_0=2.25*10^{-3}\alpha_s^6M_p^3/(M_\sigma M_\rho^4a(fm)^3)$ and
\ba
Z(\rho)=\int\frac{d^2p}{(2\pi)^2}|B(\rho,p)|^2, \nn \\
B(\rho,p)=\int\frac{d^2l}{C^2}f(l,p)exp(i\vec{l}\vec{\rho}). \ea
Numerical estimations of the cross section of one sigma meson
production for the give $\sigma_1=10 mb$.

\section{conclusion}

The results given above can be applied to explain the excess of
the positive muons compared with the negative ones produced by
cosmic ray interaction with the Earth surface. Really one can
neglect QED mechanisms of production of the charged pions  in
favor to strong interactions one. It turns out that the main
mechanism is the peripheral production of the pions pairs (with
the subsequent decay to muons) in the high energy cosmic ray
proton collisions with the nuclei of nitrogen or oxygen in the
Earth atmosphere.

Keeping in mind the atmosphere gas density one can be convinced
that at least one direct collision of cosmic ray proton with the
nuclei take place.  The QCD mechanism contribution for impact
parameters $\rho$ exceeding size of nuclei ($\rho\gg
A^{\frac{1}{3}}fm$) is also small.
 Travelling through the nuclei the cosmic proton have a direct collision with the protons
and the neutrons of the nuclei $\rho\leq 1fm$ (such kind of
collisions produce positive charged pions due to the decay of the
excited resonances) or have the peripheral collisions, when the
pairs of pions are produced ($1 fm <\rho<A^{\frac{1}{3}} fm$).

The number of positive charged pions produced in direct collisions
$N_d$, is proportional to $A^{1/3}$ with the characteristic atomic
number $A=14$. The number of pion pairs produced in the peripheral
collisions can be estimated as  \ba
 N_p=M_p^2\sigma_1\approx 7.8.
 \ea

For the ratio of the positive charged muons to the negative charged ones $R=\frac{N_{\mu+}}{N_{\mu-}}$ we have
\ba
R_{th}=1+\frac{N_d}{N_p}=1.32.
\ea
 This quantity can be compared with the recent experimental value \cite{MINOS} (here only hard muons are taken into account).
 \ba
 R_{exp}=\frac{N_{\mu^+}}{N_{\mu_-}}=1.4\pm 0.003.
 \ea
These values are in reasonable agreement.

\section{Acknowledgement}
We are grateful to E. Kokoulina for taking part at the initial state of this work.
We are grateful to A. Kotinjan  and O. Kuznetsov for discussing the possibility to study the Pomeron physics
at COMPASS facility with pion beams and to R. Peschanskii for discussions. We are grateful
to a grants INTAS 05-1000-008-8323; MK-2952.2006.2
for financial support.


\end{document}